\documentclass[aps,twocolumn,showpacs,groupedaddress,amssymb,10pt]{revtex4}

\usepackage{epsfig}
\usepackage{color}

\def\etal{et al.}

\begin{document}

\title{Magnetic Sublevel Population and Alignment for the Excitation of H- and He-like Uranium
in Relativistic Collisions}

\author{
A.~\surname{Gumberidze}$^{1,2}$, S.~\surname{Fritzsche}$^{2,3,4}$,
S.~\surname{Hagmann}$^{3,7}$, C.~\surname{Kozhuharov}$^{3}$,
X.~\surname{Ma}$^{8}$, M.~\surname{Steck}$^{3}$,
A.~\surname{Surzhykov}$^{3,5}$, A.~\surname{Warczak}$^{6}$,
Th.~\surname{St\"ohlker}$^{3,5,9}$} \vspace{5cm}

 \affiliation{$^1$ExtreMe Matter Institute EMMI and Research Division, GSI Helmholtzzentrum f\"{u}r Schwerionenforschung, D-64291 Darmstadt, Germany \\
$^2$FIAS Frankfurt Institute for Advanced Studies, D-60438
Frankfurt am Main, Germany \\
$^3$GSI Helmholtzzentrum f\"ur Schwerionenforschung, D-64291 Darmstadt, Germany\\
$^4$Department of Physics, P.O.~Box 3000, Fin-90014 University of
Oulu, Finland \\
$^5$Physikalisches Institut, Ruprecht-Karls-Universit{\"a}t
Heidelberg, Germany \\
$^6$Institute of Physics, Jagiellonian University,
Krakow, Poland \\
$^7$Institut f\"ur Kernphysik University of Frankfurt 60486 Frankfurt Germany\\
$^8$Institute of Modern Physics 730000 Lanzhou China\\
$^9$Helmholtz-Institut Jena, D-07743 Jena, Germany }

\begin{abstract}
We have measured the alignment of the L-shell magnetic-substates
following the K-shell excitation of hydrogen- and helium-like
uranium in relativistic collisions with a low-Z gaseous target.
Within this experiment the population distribution for the L-shell
magnetic sublevels has been obtained via an angular differential
study of the decay photons associated with the subsequent
de-excitation process. The results show  a very distinctive
behavior for the H- and He-like heavy systems. In particular for
$K \rightarrow L$ excitation of He-like uranium,  a considerable
alignment of the L-shell levels was observed. A comparison of our
experimental findings with recent rigorous relativistic
predictions provides a good qualitative and a reasonable
quantitative agreement, emphasizing the importance of the
magnetic-interaction and many-body effects in the strong-field
domain of high-Z ions.
\end{abstract}



\maketitle

\section{Introduction}

Relativistic collisions involving heavy high-Z ions provide an
opportunity for comprehensive testing of our understanding of
elementary processes related to ultrafast electromagnetic
interactions and serve thus as an important testing ground for
fundamental atomic theories. One of the elementary processes is
the ionization of a strongly bound projectile electron caused by
the Coulomb interaction with the target nucleus. For relativistic
\cite{Eichler/Meyerhof:95} and even ultra-relativistic collisions
\cite{Belkacem1}, this process has been the subject of intense
experimental and theoretical studies over many years where the
validity of pertubation theory, the proper choice of wavefunction
and the relevance of the magnetic part of the interaction were in
the focus of the investigations (see
\cite{Eichler/Meyerhof:95,Rymuza93,Belkacem1,Belkacem2,Krause,Cla97,Kra98,Mom96,Stoe97,Ion99,Fritzsche2001}
and references therein). Very recently K-shell ionization has
attracted particular attention as a tool for atomic structure
investigation \cite{Trot2010} since for few-electron ions this
process exhibits an extra ordinary state selectivity
\cite{Rad2006}. In general, an overall agreement between
experimental data and refined theoretical treatments can be stated
but there is still an indication of a small but systematic
deviation between experiment and theory for very asymmetric
collision systems \cite{Rymuza93,Stoe97} where first order
perturbation theory is expected to be an excellent approximation.

Compared to ionization, Coulomb excitation is mediated by the same
interaction mechanism, but, the projectile electron is excited
into a bound state of the ion and not into the continuum.
Therefore, a much better experimental control can be expected by
measuring the de-excitation photons. Earlier experimental and
theoretical studies for high-Z have focused on total
cross-sections for K-shell electron excitation by using H- and
He-like Bi and H-like Au projectiles
\cite{Stoe98,Stoe98a,IoS01,gumb10}. The results obtained have
shown, in particular, the importance of the magnetic part of the
Lienard-Wiechert interaction, namely the need to include the
magnetic term coherently into the excitation amplitude, leading to
reduced total excitation cross sections. This coherent
incorporation of the electric and magnetic parts of the
interaction potential is in contrast to any quasi-relativistic
approach in which these contributions are added incoherently and
which has been applied quite successfully for the description of
the ionization process. In Ref. \cite{Ludziejewski00},
simultaneous excitation and ionization of He-like uranium has been
addressed. The obtained experimental cross sections when compared
to relativistic calculations based on the independent particle
approximation and first-order perturbation theory have provided
reasonable agreement for lighter targets whereas for heavy targets
systematic deviations have been observed. In \cite{najjari08},
relativistic symmetric eikonal model was applied for the
description of this process which provided significantly better
agreement with the experimental results.

Compared to total cross-sections, differential measurements often
provide more detailed insights into the mechanism of a particular
process \cite{Ma01}. As an example, for radiative electron capture
(REC) into high-Z ions, former investigations of angular
differential photon emission and alignment of the associated
excited states gave access to many subtle details of relativistic
atomic collision dynamics as well as of the  electronic structure
of high-Z ions \cite{Eichler:07,Stoe97a,Surzhykov02}. In addition,
we like to note that the alignment and polarization phenomena in
ion-atom collisions have been studied in several earlier
investigations in particular for low-Z ions (see for example
\cite{Pedersen75,Ellsworth79,Church82,Palinkas85}). The K-shell
x-ray radiation produced by electron capture (and in few cases via
excitation) in few-electron ions from target atoms has been found
to be anisotropic and the corresponding orientation and alignment
parameters have been obtained by measuring the x-ray polarization
or angular differential cross sections
\cite{Pedersen75,Ellsworth79,Church82,Palinkas85}. However, in
those studies mainly low-Z ions in the relatively low collision
energy regime have been investigated. Besides, in most of the
cases it has not been possible to unambiguously resolve individual
transitions contributing to the observed x-ray lines.

In this work, we present the first experimental study of the
angular differential photon emission following the K-shell
excitation of the heaviest H- and He-like systems in relativistic
collisions with N$_2$ molecules. From the angular-resolved
measurement of the x-ray emission following K-shell excitation we
were able to determine the associated population distributions for
the magnetic L-shell sub-levels belonging to the $2p_{3/2}$ state
in H-like uranium and to the $^1P_1$ and $^3P_1$ state in He-like
uranium, respectively. The results display a very different
behavior for the two (hydrogen- and helium-like) systems under
consideration, contradicting the usual viewpoint that the coupling
of the electrons and their interaction only plays a minor role in
the high-Z regime.

The paper is structured as follows; in the next section, the
experimental arrangement as well as the experimental method is
described, in section \ref{results}, we then present the data
analysis and compare our experimental results for sub-shell- and
angular-differential cross-sections with the predictions of a
fully relativistic theory. In section \ref{summary}, finally, a
short summary and conclusions are given.


\section{Experimental arrangement} \label{experiment}
Experimental information about Coulomb excitation of one- and
few-electron projectiles occurring in relativistic atomic
collisions is very scarce. The lack of data arises mainly from
experimental difficulties due to the fact that excitation is not
accompanied by a projectile charge exchange. As a consequence,
this process can usually only be studied in single pass
experiments by measuring the photon production in coincidence with
primary beams of low intensity \cite{Stoe98,Ludziejewski00}.
Recently, an alternative experimental approach has been introduced
at the Experimental Storage Ring (ESR) at GSI where projectile
excitation has been studied by detecting the projectile x-ray
emission in anti-coincidence with charge exchange
\cite{Kraemer99,gumb10}. In our study we have utilized this
technique to extend the earlier investigations to a more detailed
analysis of the Coulomb excitation of U$^{91+}$, U$^{90+}$  and
their angular-differential x-ray emission in relativistic
collisions with N$_2$ target.

\begin{figure}
\begin{center}
\includegraphics[width=0.46\textwidth]{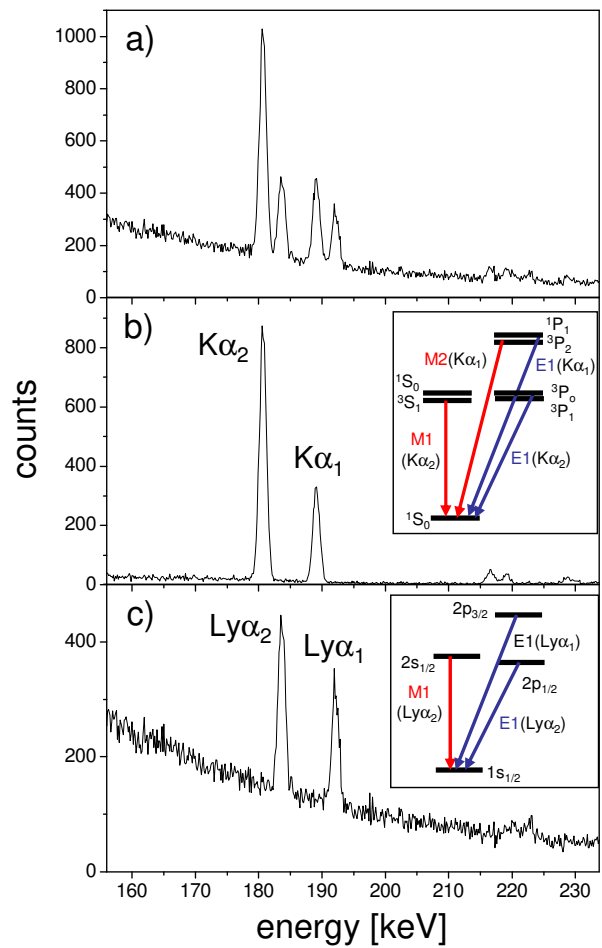}
\caption{X-ray spectra recorded for 217 MeV/u
U$^{91+}$$\rightarrow$N$_2$ collisions at the forward angle of
close to 10~$^\circ$: (a) total emission spectrum without
coincidence requirement; (b) photons in coincidence with electron
capture, $L \rightarrow K$ transitions in He-like uranium; (c)
photons in anti-coincidence with electron capture, $L \rightarrow
K$ transitions in H-like uranium. The insets show the
corresponding level schemes and transitions.}
\label{fig1}       
\end{center}
\end{figure}

The experiment was performed at the ESR by using H- and He-like
uranium ions delivered by the heavy ion synchrotron (SIS) at an
energy of 217~MeV/u. An efficient electron cooling in the ESR
storage ring provided beams with very low emittance (beam size of
less than 5 mm) and a longitudinal momentum spread of $\Delta p /p
\,\sim\, 10^{-5}$ which enabled storage of the beam with long
lifetimes as well as a decrease of the uncertainties due to the
relativistic Doppler effect. After injection in the ring and
following cooling, the ion beam interacted with a supersonic jet
of N$_2$ target. For the experiment, the atomic physics photon
detection chamber of the internal target of the ESR was utilized.
Here, projectile x-rays produced in collisions of the stored ion
beams with the jet-target were detected by an array of solid state
detectors, covering observation angles in the range between
10$^{\circ}$ and 150$^{\circ}$ with respect to the beam axis. The
Ge(i) photon detectors were energy and efficiency calibrated
before the experiment using a set of appropriate  radioactive
sources. In addition, those projectile ions that captured an
electron were detected after the next dipole magnet of the ESR
with a multiwire proportional counter (MWPC). A detailed
description of the detection setup at the ESR jet-target and of
the utilized anticoincidence technique can be found in
\cite{gumb03,gumb10} and in references therein.

As an example, we depict in Fig. \ref{fig1} x-ray spectra recorded
for U$^{91+}$$\rightarrow$ N$_2$ collisions at the forward angle
of close to 10$^{\circ}$.
In the total spectra (a), measured without any coincidence
condition, the characteristic transitions arising from both
electron capture (K$\alpha$ transitions in He-like uranium) and
from excitation (Ly$\alpha$ transitions in H-like uranium) are
clearly visible. Applying a coincidence requirement with the
down-charged projectile, we obtain the characteristic x-ray
spectra corresponding exclusively to the events of capture of one
electron from the $N_2$ into initially H-like uranium (U$^{91+}$).
Furthermore, by substraction of the spectrum corresponding to a
capture (b) from the total one (a) we obtain the spectrum in
anti-coincidence with the projectile charge exchange (c) which
comprises only the events corresponding to $K \rightarrow L$
excitation (and following decay) of the projectile electron.
Indeed, no He-like K$\alpha$ transitions are observed in the
anti-coincidence spectrum. This also proves that the MWPC detector
used for particle detection operates with a detection efficiency
very close to 100\%. The exponential background observed in the
excitation spectrum stems from electron bremsstrahlung. In the
same way, a spectrum containing only the events corresponding to
$K \rightarrow L$ excitation has also been obtained for He-like
projectiles. In this case, the situation is even more favorable
because K-shell transitions can only be produced by a K-shell
excitation.

Finally, we like to note that the data for initially H- and
He-like uranium have been acquired with an identical detector
setup in sequential order during the same beam time. The ion beam
position was defined by the target in both cases, and thus, stayed
unchanged. Therefore, the detector geometry with respect to the
beam axis was the same for the both beams. This is very important
since it allows to normalize the observed yields of x-ray
transitions in He-like ions to the line intensity of Ly-$\alpha_1$
transition which is  known to be precisely isotropic (see below).

\section{Data analysis and results} \label{results}
In the following, we focus on the angular distributions of
characteristic x-rays following the excitation of H- and He-like
projectiles. Such an angle-resolved analysis will enable us to
gain insight into a mechanism of formation of excited ionic
sub-states in relativistic ion-atom collisions. The information on
the magnetic sublevel population of the excited ion can be
directly extracted from the angular distribution of de-excitation
photons that can be written in the projectile (emitter) frame as:

\begin{equation}\label{1}
W(\theta)=A_0+A_2P_2(\cos\theta) \propto
               1+\beta_{20}(1-\frac{3}{2}\sin^2\theta)
\end{equation}
Here, $\theta$ is the angle between the direction of the
de-excitation photon and the beam direction while $P_2$ denotes
the second-order Legendre polynomial.

The emission pattern (1) is completely determined by the effective
anisotropy parameter $\beta_{20}$ whose particular form depends on
the transition under consideration. For example, for the
Ly$\alpha_1$ ($2p_{3/2} \to 1s_{1/2}$) decay in H-like ions, the
anisotropy $\beta_{20} = \mathcal{A}_2 f(E1,M2) /2$ includes not
only the alignment parameter:

\begin{eqnarray}\label{3}
{\mathcal{A}}_2=\frac{\sigma(\frac{3}{2},\pm\frac{3}{2})-
\sigma(\frac{3}{2},\pm\frac{1}{2})}{\sigma(\frac{3}{2},\pm\frac{3}{2})+
\sigma(\frac{3}{2},\pm\frac{1}{2})}
\end{eqnarray}

which can be written in terms of the partial cross sections
$\sigma(j \mu)$ for populating the magnetic substates $j \mu$, and
the so-called structure function $f(E1,M2)$ \cite{Surzhykov02}.
This function describes the interference between the leading E1
and the much weaker M2 decay channels and contributes by as much
as $f (E1,M2) = 1.28$ to the angular distribution of the
characteristic photon emission from H-like uranium ions. In
contrast to the Ly$\alpha_1$ transition, no multipole-mixing can
occur for the decay of the $[1s, 2p_{1/2}] ^3P_1$ and $[1s,
2p_{3/2}] ^1P_1$ states in He-like ions since they can only decay
via a fast E1 transition to the groundstate. For the $K\alpha_1$
($2 ^1P_1 \to 1 ^1S_0$) and $ K\alpha_2$ ($2 ^3P_1 \to 1 ^1S_0$)
transitions, therefore, the angular distribution (1) is governed
by the anisotropy parameter $\beta_{20} = \mathcal{A}_2 /\sqrt{2}$
where the alignment parameter reads:

\begin{equation}
\mathcal{A}_2 = \sqrt{2} \frac{\sigma(1, \pm 1) - \sigma(1, 0)}{
(2 \sigma(1, \pm 1) + \sigma(1, 0))}
\end{equation}
Here, the partial cross sections $\sigma(J,M)$ describe the
dynamics of excitation of the \textit{two-electron} projectiles
\cite{Surzhykov08}.

\begin{figure}
\begin{center}
\includegraphics[width=0.45\textwidth]{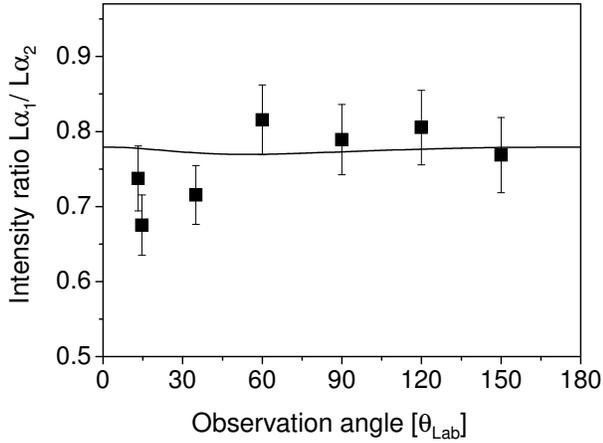}
\caption{The intensity of Ly$\alpha_1$-transition normalized to
the Ly$\alpha_2$ line intensity as function of observation angle
for 217 MeV/u U$^{91+}$$\rightarrow$N$_2$ collisions. The solid
line refers to a fit of Eq. (1) (transformed into the laboratory
frame) to the data .}
\label{fig2}       
\end{center}
\end{figure}

\begin{figure}
\begin{center}
\includegraphics[width=0.46\textwidth]{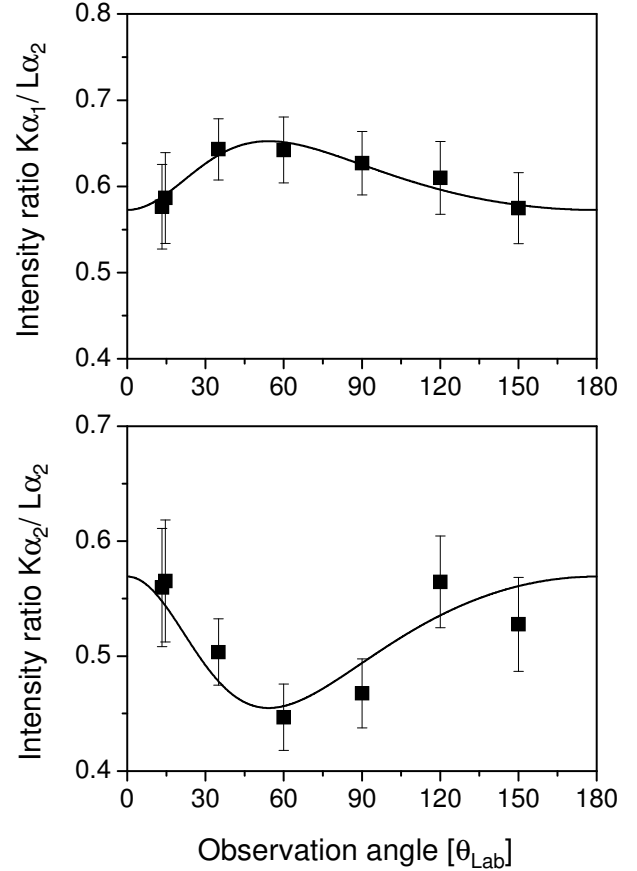}
\caption{Angular distributions of K$\alpha_1$ (top part) and
K$\alpha_2$ (bottom part) measured for K-shell excitation of
He-like uranium in collisions with N$_2$ at 217 MeV/u. In
addition, fits of Eq. (1) (transformed into the laboratory frame)
to the corresponding experimental data are shown by solid lines.}
\label{fig3}       
\end{center}
\end{figure}

\begin{figure*}
\begin{center}
\includegraphics[width=0.82\textwidth]{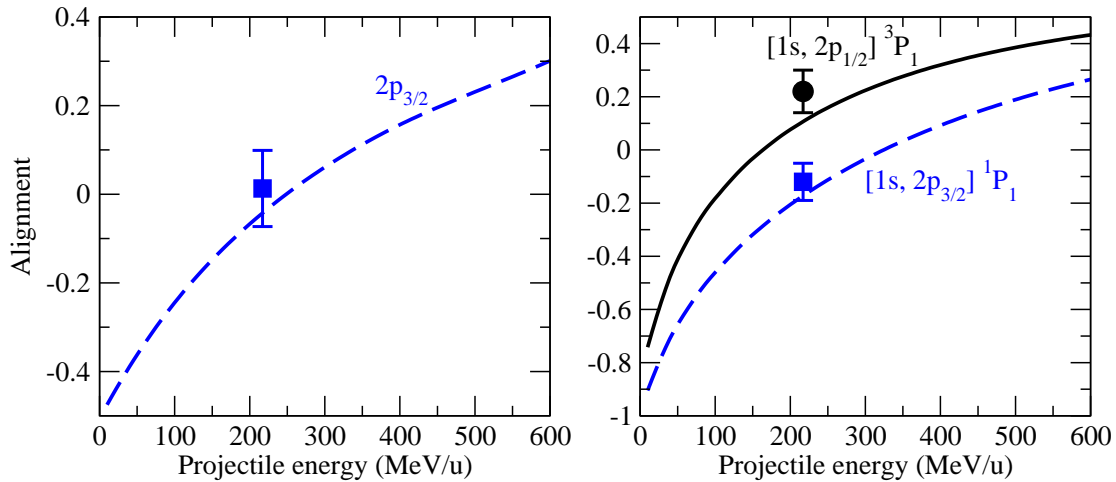}
\caption{Alignment parameters $A_2$ of the 2p$_{3/2}$ state of
hydrogen--like (left panel) and the $[1s_{1/2}, 2p_{1/2}] ^3P_1$
as well as the $[1s_{1/2}, 2p_{3/2}] ^1P_1$  states of
helium--like (right panel) uranium ions following K-shell
excitation. Results of relativistic calculations are compared with
the experimental findings for collisions with N$_2$ target at the
energy of 217 MeV/u.}
\label{fig4}       
\end{center}
\end{figure*}

In our experiment, we strongly benefit from the fact that the
Ly$\alpha_2$ transition arising from the decay  of the 2s$_{1/2}$
and 2p$_{1/2}$ levels is  known to be precisely isotropic.
Consequently, this line provides an ideal tool to measure a
possible anisotropy of the close-spaced Ly$\alpha_1$ or K$\alpha$
transitions. By using this transition for normalization purposes,
various systematic effects, associated for example with solid
angle corrections, possible error in detector efficiency
calibration, etc. cancel out. This technique has been successfully
applied in previous studies \cite{Stoe97a}. In Fig. \ref{fig2},
the result for the emission pattern of the Ly$\alpha_1$
(2p$_{3/2}\rightarrow$1s$_{1/2}$)  transition is shown, normalized
to the Ly$\alpha_2$ intensity. The angular distribution is fairly
isotropic. From this experimental result we extract the value of
the alignment parameter for the 2p$_{3/2}$ state by fitting the
eq. \ref{1} (transformed into the laboratory frame) to the
observed angular distributions (see Fig. \ref{fig2}). The
anisotropy coefficient $\beta_{20}$ and an overall amplitude of
the fit function were kept as free fit parameters. From this
procedure we obtained a value of $0.013\pm0.086$ for the alignment
 $\mathcal{A}$$_2$ that is consistent with the
isotropic angular distribution. The quoted uncertainty is entirely
of statistical origin. This finding is in agreement with
theoretical predictions (see Fig. \ref{fig4}) showing no alignment
for the particular collision energy used in the experiment. Here,
we like to note that this result is markedly different to earlier
findings obtained for the same transition in $U^{91+}$
(2p$_{3/2}\rightarrow$1s$_{1/2}$) but caused by REC (the same beam
energy and the same  target) \cite{Stoe97a,Surzhykov02} where a
very strong alignment has been observed. Moreover, in our
experiment we observed a very different behavior for the K-shell
excitation of He-like uranium as compared to the H-like case.
In Fig. \ref{fig3}, the K$\alpha_1$ and K$\alpha_2$ the angular
distributions as measured for K-shell excitation of He-like
uranium in 217 MeV/u U$^{90+}$$\rightarrow$N$_2$ collisions are
presented, normalized to the Ly$\alpha_2$ yield as obtained for
H-like uranium. As seen from the figure, the data (especially
K$\alpha_2$ ($[1s_{1/2},2p_{1/2}]^3P_1\rightarrow 1^1S_0$))
exhibit a pronounced deviation from a constant intensity ratio.
Let us note here, moreover, that for the case of K-shell
excitation of high-Z He-like ions, only the
$[1s_{1/2},2p_{3/2}]^1P_1$ and the $[1s_{1/2},2p_{1/2}]^3P_1$
states are predicted to be populated and contribute to the
observed K$\alpha_1$ and K$\alpha_2$ transitions
\cite{Surzhykov08}. In both cases an alignment of the different
sublevels is possible. As it is seen from Fig. \ref{fig3}, there
is a significant positive alignment for the
$[1s_{1/2},2p_{1/2}]^3P_1$ level (K$\alpha_2$ transition) and a
relatively weak negative alignment for the
$[1s_{1/2},2p_{1/2}]^1P_1$ state (K$\alpha_1$ transition).
Applying the same procedure as for the case of the H-like ions
(described above) we obtained values of $-0.12\pm0.07$ and
$0.22\pm0.08$ for the alignment parameters of the $^1P_1$ and
$^3P_1$ states, respectively. The quoted uncertainties are
entirely of statistical origin.

We can compare these experimental results with values of $-0.172$
and $+0.110$ given by a recent fully relativistic calculations
\cite{Surzhykov08,Fritzsche01}. In these computations, both the
coupling of the electrons as well as their repulsive interaction
have been taken into account within the framework of the
multiconfiguration Dirac-Fock method (MCDF). These calculations
included moreover the cascade feeding (following an initial
excitation into states with $n \geq 3$) and its effect upon the
aligment of the $^1P_1$ and $^3P_1$ states, in addition to the
aligment as obtained for a direct $K \rightarrow L$ excitation. As
known for the photo-induced excitation of atoms and ions
\cite{Kabachnik07}, the different sign in the alignment of the
$^1P_1$ and $^3P_1$ states arises from the different coupling of
the two electrons. Though the aligment differs quite considerably
for these two states, it may be both negative (low projectile
energies) or positive (high projectile energies) as seen from
figure \ref{fig4}. Let us note that this qualitative behavior of
the alignment as a function of energy has been already predicted
by earlier non-relativistic calculations for $^1S_1 \rightarrow
^1P_1$ excitation \cite{VanDenBos67,VanDenBos69,Bell84} and it
holds also true for excitation of H-like systems ($1s_{1/2}
\rightarrow 2p_{3/2}$) \cite{Stoe98}. This finding is simply
related to the fact that with increasing energy more angular
momentum is transferred to the system so that transitions with
$\Delta m$=$\pm 1$ (He-like system) start to dominate
\cite{Stoe98,IoS01,Surzhykov08}. From the comparison in figure
\ref{fig4}, a very good agreement with the experimental result for
the alignment of the $^1P_1$ state can be seen. For the $^3P_1$
level, the theoretical value is smaller than the experimental one,
however, the deviation is not particularly pronounced due to the
experimental uncertainty. Here it is important to note that the
theoretical values include only the Coulomb excitation due to the
target nucleus and omit completely the process of electron impact
excitation. Coulomb excitation caused by the nuclear charge $Z_T$
of the target scales with $Z_T^2$ whereas the cross section for
electron impact excitation scales linearly with the amount of
target electrons available. Therefore, for the particular case of
nitrogen target we might expect that electron impact excitation
contributes by about fifteen percent to the overall excitation
cross section.


\section{Summary and Outlook} \label{summary}
In summary, we have performed the first experimental study of the
magnetic-sublevel population for the K-shell excitation of
hydrogen- and helium-like uranium in relativistic collisions with
a low-Z gaseous target. The information about the population of
the magnetic sublevels in this process has been obtained via an
angular differential study of the decay photons associated with
Coulomb excitation. The results presented in this work show a
markedly different behavior for the different ion species (H-like
or He-like). The Lyman transition
(2p$_{3/2}\rightarrow$1s$_{1/2}$) following K-shell excitation of
H-like uranium has been found to be nearly isotropic and therefore
the population of the magnetic sublevels follows a statistical
distribution. For the $K \rightarrow L$ excitation of He-like
uranium, in contrast, we have measured an non-zero alignment for
both $^3P_1$ and $^1P_1$ states but with alignment parameters of
opposite sign. Though perhaps not very surprising in its own, this
result cannot be understood just within a single-electron model
but requires to account for the coupling of the electrons as well
as their interaction. This is remarkable as we are dealing with
high-Z ions for which  the interaction among the electrons is
commonly assumed to be of minor importance. The experimental data
agree well with recent theoretical predictions \cite{Surzhykov08}
in which both, the relativistic and magnetic interaction effects
were taken into account  and, thus, provide a meaningful
description of the K-shell excitation process in relativistic
collisions of high-Z ions with low-Z targets. However, since this
is the only measurement for He-like uranium up to date, additional
studies with different targets and different collision energies
would be desirable to unravel especially the role of electron
impact excitation in collisions of high-Z ions with gaseous
targets.
 These investigations will additionally be supported by
the observation of linear polarization of the de-excitation x-rays
as very recently introduced in experiments at the ESR
\cite{Weber2010}

\newcommand{\pppp}[4]{{\it{#1}}~{\bf{#2}}~(19{#4})~{#3} .}
\newcommand{\ppp}[3]{{\bf{#1}}~(19{#3})~{#2} .}

\section{Acknowledgment}

This work was supported by the Helmholtz Alliance Program of the
Helmholtz Association, contract HA216/EMMI "Extremes of Density
and Temperature: Cosmic Matter in the Laboratory".
A.S.~acknowledges support from the Helmholtz Gemeinschaft and GSI
under the project VH-NG-421 and S.F.~those of the FiDiPro program
of the Finnish Academy.

\end{document}